\documentclass[%
 aip,
 jap,
 amsmath,amssymb,
reprint,%
]{revtex4-1}

\usepackage{graphicx}
\usepackage{dcolumn}
\usepackage{bm}

\begin{document}

\preprint{AIP/JAP-1}

\title[Helical edge magnetoplasmon]{Helical edge magnetoplasmon in the quantum Hall effect regime}

\author{Sanderson Silva}
\author{O. G. Balev}
\email{ogbalev@ufam.edu.br}
\affiliation{Departamento de F\'{\i}sica, Universidade Federal do Amazonas, 69077-000,
Manaus, Amazonas, Brazil}


\date{\today}

\begin{abstract}
 We present the microscopic treatment of edge magnetoplasmons
(EMPs) for the regime of not-too-low temperatures defined by the
condition $\hbar \omega _{c}\gg k_{B}T\gg \hbar v_{g}/2\ell _{0}$,
where $v_{g}$ is the group velocity of the edge states, $\ell
_{0}=\sqrt{\hbar /m^{\ast }\omega _{c}}$ is the magnetic length
and $\omega _{c}$ is the cyclotron frequency. We find a weakly
damped symmetric mode, named helical edge magnetoplasmon, which is localized at
the edge states region for filling factors $\nu =1, 2$ and
\textit{very strong dissipation} $\eta _{T}=\xi /k_{x}\ell
_{T}\agt\ln (1/k_{x}\ell _{T})\gg 1$, where the characteristic
length $\ell _{T}=k_{B}T\ell _{0}^{2}/\hbar v_{g}\gg \ell _{0}/2$
with $\xi $ being the ratio of the local transverse
conductivity to the local Hall conductivity at the edge states
and $k_{x}$ is the wave vector along the edge; here 
other EMP modes are strongly damped. The
spatial structure of the helical edge magnetoplasmon, transverse to the edge, is
strongly modified as the wave propagates along the edge. In the
regime of \textit{weak dissipation}, $\eta _{T}\ll 1$, we obtain
exactly the damping of the fundamental mode as a function of
$k_{x}$. For $\nu =4$ and weak dissipation we find that the
fundamental modes of $n=0$ and $n=1$ Landau levels (LLs) are
strongly renormalized due to the Coulomb coupling. 
Renormalization of all these EMPs coming from a metal
gate and air half-space is studied.
\end{abstract}


\maketitle   

\section{Introduction}


Edge magnetoplasmons (EMPs) in the two-dimensional electron system (2DES)
are chiral collective excitations propagating along the edge of the 2DES in
the presence of a normal magnetic field $B$. They have received much
attention after pioneering works in the 80's.\cite{allen83,mast85,glattli85}
Many experimental studies have been performed for confined electrons in
AlGaAs-GaAs heterostructures \cite%
{allen83,galchenkov86,talyanskii89,wassermeier90,grodnensky91,ashoori92,%
talyanskii92,zhitenev93,talyanskii95,ernst96,deviatov97,balaban97,sukhodub04,%
kukushkin04,kukushkin05,dorozhkin05}
and for surface electrons on liquid helium.\cite%
{mast85,glattli85,kirichek95} The advent of time-resolved techniques\cite%
{ashoori92,zhitenev93,ernst96,sukhodub04} and recent experimental
techniques based on $B$-periodic oscillations in the photovoltage
and in the longitudinal resistance of 2DES, induced by microwave
radiation and attributed to the
interference of coherently excited EMPs, \cite%
{kukushkin04,kukushkin05,dorozhkin05} increases greatly the
interest in EMPs studies. EMPs are sensitive to reconstruction of quantum Hall edges%
\cite{wen91,melikidze04} and can probe the interactions properties in a
quantum Hall line junction system.\cite{papa05}
A lot of theoretical works have investigated
the characteristics of EMPs and
different edge-wave mechanisms are proposed or
used.\cite{fetter86,volkov88,wen91,stone91,aleiner94,chamon94,%
giovanazzi94,han97,zulicke97,balev97,balev98,balev99,balev00,%
hansson00,johnson03,melikidze04,papa05,mikhailov06}

Theoretically EMPs often are studied
employing models entirely classical, see \textit{e.g.}
Refs.\cite{volkov88,aleiner94}. In these edge-wave mechanisms the
charge density varies at the 2DES edge, but the position of the
2DES edge is kept constant. Different, quantum-mechanical
edge-wave mechanisms are developed in Refs. \cite{wen91,stone91,chamon94,giovanazzi94}. 
In these models only the edge of 2DES
is varying while the density profile, with respect to the fluctuating edge,
coincides with that of the unperturbed 2DES. Recently, a microscopic model
has been proposed in Refs.\cite{balev97,balev98} that effectively incorporates
both above edge-wave mechanisms in the quantum Hall effect (QHE) regime.

The study of the EMPs with this new edge-wave mechanism is elaborated
in Refs. \cite{balev97,balev98,balev99} in the limit of low temperatures
given by $k_{B}T\ll \hbar v_{g}/2\ell _{0}$, where $v_{g}$ is the group
velocity of the edge states and the magnetic length $\ell _{0}=\sqrt{\hbar
/m^{\ast }\omega _{c}}$ with the cyclotron frequency $\omega
_{c}=|e|B/m^{\ast }c$. Later, the approach of Refs. \cite{balev97,balev98}
has been extended in Ref. \cite{balev00} for not-too-low temperatures such that $%
\hbar \omega _{c}\gg k_{B}T\gg \hbar v_{g}/2\ell _{0}$, where the
typical scale of in-plane components of the wave electric field
$\bf{E}$ is of the order of the characteristic length $\ell
_{T}=\ell _{0}^{2}k_{B}T/\hbar v_{g}\gg \ell _{0}/2$. In Ref.
\cite{balev00} it is shown that the main contribution to the
damping of EMPs is given by dissipation effects localized within a
narrow region of width of the order of $\ell _{T}$, nearby the
edge states. For $\nu =1, 2$, weakly damped EMPs are obtained in
Ref. \cite{balev00} only in the \textit{weak dissipation} regime
$\eta _{T}\ll 1$ or in the \textit{strong dissipation} regime
defined by $\ln (1/k_{x}\ell _{T})\gg \eta _{T}\gtrsim 1$ and for
homogeneous samples (i.e., without gate or air region close to the
2DES). 

In the present paper the microscopic approach of Ref.
\cite{balev00} is extended to study EMPs at \textit{not-too-low
temperatures} both in homogeneous samples and in samples with a
metal gate or air on the top of the sample at distance $d$ from
the 2DES. We study, at $\nu =1, 2$, EMPs for very strong
dissipation at the edge, \textit{i.e.}, for $\eta _{T}\gtrsim \ln
(1/k_{x}\ell _{T})\gg 1$, and a new weakly damped mode, which we
call \textit{helical edge magnetoplasmon} or 
\textit{edge helicon}, is obtained. In particular, we obtain that the real
part of helical edge magnetoplasmon frequency $Re\omega $ for the
homogeneous sample and very strong dissipation almost
coincides with that of the fundamental EMP, obtained in Ref.
\cite{balev00} for $\eta _{T}\ll 1$. This result is rather
unexpected because, in particular, all other EMPs (calculated from
the same secular determinant as the helical edge magnetoplasmon mode)
are now strongly damped, with $%
Re\omega /(-Im\omega )\sim 1/\eta _{T}\ll 1$. We
show that the spatial structure of the helical edge magnetoplasmon, transverse to the
edge, is strongly modified as the wave propagates along the edge.
In addition, for the gated sample and \textit{weak dissipation},
$\eta _{T}\ll 1$, the acoustic-like dispersion for the fundamental
mode is found for some ranges of $k_{x}$ and comparison with
pertinent qualitative estimations of Ref. \cite{zhitenev95} is
given. Finally, we study the fundamental modes of $n=0$ and $n=1$
LLs and their strong renormalization due to the Coulomb coupling
for $\nu =4$. and weak dissipation, $\eta _{T}\ll 1$,

We emphasize three essential points that we take into account in our
approach in order to determine the dispersion relation as well the spatial
structures of the EMPs. The first one is the role of dissipation that 
is mainly localized at the edge states when a smooth lateral
confinement is assumed and $\hbar \omega _{c}/k_{B}T\gg 1$ \cite{balev93}; 
it is related with the scattering by acoustic phonons and becomes 
exponentially small for $v_{g} < s$ in comparison with 
the case $v_{g} > s$, where $s$ is the speed of sound.
The second one is the\ very realistic form of the unperturbed electron
density $n_{0}(y)$ across the edge. It was shown\cite{balev00} that for
not-too-low temperatures and $\nu =1, 2$ the density profile $n_{0}(y)$ is
well approximated by $n_{0}(y)/n_{0}=[1-\tanh (Y/2)]/2$, where $%
Y=(y-y_{re})/\ell _{T}$ with $y_{re}$ being the edge of the $n=0$ LL when
the condition $k_{e}\gg k_{B}T/\hbar v_{g}\gg 1/2\ell _{0}$ is fulfilled.
Here the characteristics edge wave number $k_{e}=(\omega _{c}/\hbar \Omega )%
\sqrt{2m^{\ast }\Delta _{F}}$, where $\Delta _{F}$ is the Fermi
energy measured from the bottom of the $n=0$ LL, i.e., $\Delta
_{F}=E_{F0}-\hbar \omega _{c}/2$; $n_{0}$ is the density in the
interior part of the channel and $E_{F0}$ is the Fermi level
measured from the bottom of the lowest electric subband. In
addition, we show here that the density profile is more
complicated for $\nu =4$ (see the solid curve in Fig. 8): the steep drop of $%
n_{0}(y)$ at the edge is splitted in two steps - one is centered at the $n=0$
LL edge, $y_{re}$, and the other step is centered at the $n=1$ LL edge, $%
y_{re}^{(1)}$.

The third point is related with effects of the strong
electron-electron interaction in the EMPs. In particular, at $\nu
=1, 2$, the weakly damped helical edge magnetoplasmon mode is manifested, 
for the regime of very strong dissipation, only due to the strong
self-consistent electron-electron interaction. Indeed, if we
assume that the electron-electron interaction is weak then this
mode becomes strongly damped as any other EMP mode. In particular,
by neglecting the coupling of the monopole term (the main term for
the helical edge magnetoplasmon) with any multipole term we obtain for
the homogeneous sample and $k_{x}\ell_{T} \ll 1$, according to Ref.
\cite{balev00}, that $Re \omega /(-Im\omega )\approx
[4\ln(1/k_{x}\ell_{T})]/(3\eta _{T}) \ll 1$ for typical conditions
of our Fig. 2. I.e., in this crude approximation the mode is
strongly damped. However, the exact results for the helical edge magnetoplasmon
presented in Fig. 2 shows that $Re \omega
/(-Im\omega ) > 17$. Point out, very close gate also can
essentially suppress electron-electron interaction within 2DES.
Further, for $\nu =4$, electron-electron interaction leads to
strong Coulomb coupling of the fundamental EMP of $n=0$ LL with
the fundamental EMP of $n=1$ LL, resulting in strong
renormalization of their dispersion relations.

We show here that the combination of these factors strongly
modifies properties of the EMPs. In addition, it deserves
to be pointed out that a many-body study of LLs spectra, at $\nu
=1$, nearby the edge states show\cite{irina2001} that electron correlations can
essentially facilitate the appearance of not-too-low temperatures
regime (in particular, for $T\sim 1$K):  as the renormalized
group velocity $v_{g}$ can become\cite{irina2001} $\propto T$ for $4.2$K$\agt T \agt 0.3$K. 

The organization of the paper is as follows. In Sec. II we present briefly 
the general formalism \cite{balev00} for $\nu =1, 2$ with 
the modifications that appear due to the gate or the air half-space. In Sec. III
A we study the integral equations for symmetric and antisymmetric
EMPs for $\eta _{T}\gg 1$ and obtain the frequency spectrum and
the spatial behavior of the helical edge magnetoplasmon. We consider also the
effect of the gate and the air half-space. In Sec. III B we treat
symmetric and antisymmetric EMPs in the weak dissipation regime,
$\eta _{T}\ll 1$. In Sec. IV, we study the fundamental EMPs of
$n=0$ LL and $n=1$ LL and their coupling by Coulomb interaction
for $\nu =4$ and in the weak dissipation regime. Finally, in Sec.
V, we make the concluding remarks.

\section{Integral equations for EMPs at $\protect\nu =1, 2$}

Our model consists of a 2DES with width $W$, length $L_{x}=L$, and zero thickness, in the presence of a strong $B$
parallel to the $z$ axis. The lateral confining potential $V_{y}^{^{\prime}}$ of the 2DES, 
semi-parabolic at the left and right edges of the channel, is given as 
\begin{eqnarray}
&&V_{y}^{^{\prime}}=0, \; \text{for} \; y_{l}<y<y_{r},  \nonumber \\
&&V_{y}^{^{\prime }}=\frac{m^{\ast }\Omega^{2}}{2}(y-y_{r})^{2}, \text{for} \;y>y_{r}>0, \nonumber \\
&&\text{and} \; V_{y}^{^{\prime }}=\frac{m^{\ast}\Omega ^{2}}{2}(y-y_{l})^{2}, \; \text{for} \; y<y_{l}<0 .
\label{A0}  
\end{eqnarray}
We assume that the confinement
is smooth on the scale of $\ell _{0}$ (i.e., $\Omega \ll \omega _{c}$) and
$|k_{x}|W\gg 1$. Then it is possible to consider an EMP \cite{wave} $A(\omega
,k_{x},y)\exp [-i(\omega t-k_{x}x)]$ only along, e.g., the right edge of the
channel. 
For $\nu =1$, we assume that the spin-splitting, caused by
many-body effects, is strong enough to neglect the contribution from the
upper spin-split LL. For $\nu =2$ and $4$, we neglect spin-splitting. To
simplify notations, we omit superscript or subscript $0$ in values pertinent
to $n=0$ LL like $v_{g0}\equiv v_{g}$, $\Delta _{F0}\equiv \Delta_{F}$, $%
k_{e}^{(0)}\equiv k_{e}$, etc. For definiteness, the 2DES is considered in GaAs based
sample.

At the $\nu=1, 2$ QHE regime, with $\hbar \omega _{c}\gg k_{B}T\gg \hbar v_{g}/\ell _{0}$, 
for $\omega \ll \omega_{c}$ the 
current density components induced by an EMP have both quasistatic and local 
(due to $\min\{1/k_{x}, \ell_{T}\} \gg \ell_{0}$) form\cite{balev00} 

\begin{eqnarray}
&&j_{x}(y)=\sigma _{yy}(y)E_{x}(y)-\sigma _{yx}^{0}(y)E_{y}(y)+v_{g}\rho (\omega ,k_{x},y) , \nonumber \\
&& j_{y}(y)=\sigma _{yy}(y)E_{y}(y)+\sigma _{yx}^{0}(y)E_{x}(y),  
\label{A1}  
\end{eqnarray}
where in the wave current density, $j_{\mu}(y)$, and the wave electric field, $E_{\mu}(y)$,
the wave factor $\exp [-i(\omega t-k_{x}x)]$, and the arguments $\omega$, $k_{x}$ 
are suppressed; $\rho (\omega ,k_{x},y)$ is the wave electron charge density and
$\sigma_{\mu\gamma}$ are the components of the conductivity tensor. Here, cf. with Refs. \cite{balev93,balev00},
$\sigma _{yy}(y)=\sigma_{xx}(y) \approx \tilde{\sigma}_{yy}R_{0}(\bar{y})$, with
$R_{0}(\bar{y})=(4\ell _{T})^{-1}\cosh ^{-2}(\bar{y}/2\ell _{T})$ and 
$\bar{y}=y-y_{re}$, is defined by nonelastic scattering within the edge states region on acoustic phonons.
Notice, additional study justifies  approximation of the diagonal conductivity 
only by its dissipative (or real) part; so
$\sigma _{yy}(y)$ is strongly localized within a distance $\alt \ell _{T}$ from 
the edge states (the intersection of LL with the Fermi level). 
For GaAs based sample the main contribution to $\tilde{\sigma}_{yy}$ is due to piezoelectric acoustic(PA)-phonons;
in particular, for $v_{g}\geq s$ and $\nu =1$ we have that  
$\tilde{\sigma}_{yy}=e^{2}\ell _{0}^{2}c^{^{\prime }}k_{B}T/4\pi ^{2}\hbar ^{4}v_{g}^{3}$,
where the electron-phonon coupling constant $c^{\prime }=\hbar (eh_{14})^{2}/2\rho _{V}s$, with 
$h_{14}=1.2\times 10^{7}$V/cm, $\rho _{V}=5.31$g/cm$^{3}$, and 
$s=2.5\times 10^{5}$cm/sec; $\epsilon=12.5$. Hereafter we are using these
parameters in numerical estimates and results shown in the figures.

For $\nu =1, 2$ we have 
\begin{equation}
\sigma _{yx}^{0}(y)\approx \sigma_{yx}^{0}
[1+e^{(E_{0}(y)-E_{F0})/k_{B}T}]^{-1},  \label{A2}
\end{equation}
where $E_{n}(y)=\hbar \omega _{c}(n+1/2)+V_{y}^{^{\prime}}$, $\sigma_{yx}^{0}=e^{2} \nu/2\pi \hbar$ is the value
of the local Hall conductivity $\sigma _{yx}^{0}(y)$ in the interior part of the channel; $\sigma _{yx}^{0}(y_{re})=\sigma_{yx}^{0}/2$. 
For $k_{e}\gg k_{B}T/\hbar v_{g}\gg 1/2\ell _{0}$
we obtain \cite{balev00} that $d\sigma_{yx}^{0}(y)/dy \approx -(e^{2} \nu/2\pi \hbar )R_{0}(\bar{y})$;
the group velocity $v_{g}=(\ell_{0}^{2}/\hbar)dE_{0}(y_{re})/dy=\hbar \Omega ^{2}k_{e}/m^{\ast
}\omega _{c}^{2}$. Then using Eqs. (\ref{A1}), (\ref{A2}), the Poisson equation, and the linearized continuity equation 
we obtain the integral equation for the wave electron charge density as
\begin{eqnarray}
&&-i(\omega -k_{x}v_{g})\rho (\omega ,k_{x},y)+ \frac{2}{\epsilon}
\{k_{x}^{2}\sigma_{yy}(y)- ik_{x}\frac{d}{dy}[\sigma _{yx}^{0}(y)]
\nonumber
\\
&&-\sigma _{yy}(y)\frac{d^{2}}{dy^{2}}-
\frac{d}{dy}[\sigma _{yy}(y)]\frac{d}{dy}\}
\int_{-\infty }^{\infty }dy^{\prime }[K_{0}(|k_{x}||y-y^{\prime }|)
\nonumber
\\
&&+\beta K_{0}(|k_{x}|\sqrt{(y-y^{\prime })^{2}+4d^{2}})]
\rho(\omega ,k_{x},y^{\prime })=0,  
\label{A3}
\end{eqnarray}
where $K_{0}(x)$ is the modified Bessel function.
For $\beta=0$, Eq. (\ref{A3}) corresponds to a sample with
the homogeneous background dielectric constant $\epsilon$, whereas for
$\beta =-1$ or $\beta=\left( \epsilon -1\right) /\left( \epsilon +1\right)$
it corresponds to a sample with the metallic gate or air half-plane at a distance $d$ from the 2DES. 

Any solution $\rho (\omega ,k_{x},y) \equiv \rho(\omega ,k_{x},\bar{y})$ of  Eq. (\ref{A3}) is 
either symmetric or antisymmetric with respect to the change $\bar{y} \to -\bar{y}$; i.e., 
with respect to the right edge, $y_{re}$. Then
straightforwardly it follows from Eq. (\ref{A3}) two integral equations. One for symmetric EMP modes, with 
$\rho ^{s}(\omega ,k_{x},\bar{y})=\rho ^{s}(\omega ,k_{x},-\bar{y})$,
and other for antisymmetric EMP modes, with $\rho ^{a}(\omega ,k_{x},\bar{y})=-\rho ^{a}(\omega ,k_{x},-\bar{y})$, as
\begin{eqnarray}
&&(\omega -k_{x}v_{g})\rho ^{s,a}(\omega ,k_{x},\bar{y})-\frac{2}{\epsilon }
\{(k_{x}\sigma _{yx}^{0}-ik_{x}^{2}\tilde{\sigma}_{yy})R_{0}(\bar{y})  \nonumber
\\
&&+i\tilde{\sigma}_{yy}\frac{d}{d\bar{y}}[R_{0}(\bar{y})\frac{d}{d\bar{y}}
]\}\int_{0}^{\infty }d\bar{y}^{\prime }\{[K_{0}(|k_{x}||\bar{y}-\bar{y}
^{\prime }|)  \nonumber \\
&&+\beta K_{0}(|k_{x}|\sqrt{(\bar{y}-\bar{y}^{\prime })^{2}+4d^{2}})]\pm
\lbrack K_{0}(|k_{x}||\bar{y}+\bar{y}^{\prime }|)  \nonumber \\
&&+\beta K_{0}(|k_{x}|\sqrt{(\bar{y}+\bar{y}^{\prime })^{2}+4d^{2}})]\}\rho
^{s,a}(\omega ,k_{x},\bar{y}^{\prime })=0,  
\label{1}
\end{eqnarray}%
where and hereafter the upper (lower) sign corresponds to symmetric
(antisymmetric) EMPs.

Solutions of Eq.(\ref{1}), for $\bar{y}\geq 0$, are given as
\begin{equation}
\rho ^{s,a}(\omega ,k_{x},\bar{y})=\tilde{R}_{0}(Y)e^{-Y}\sum_{n=0}^{\infty
}\rho _{n}^{s,a}(\omega ,k_{x})L_{n}(Y),  \label{2}
\end{equation}%
where $Y=\bar{y}/\ell _{T}$, $L_{n}(Y)$ is the Laguerre polynomial, $\tilde{R%
}_{0}(Y)=(4\ell _{T})^{-1}\exp (Y)\;\cosh ^{-2}(Y/2)$. For $\bar{y}\leq 0,$
the expression for $\rho ^{s}(\omega ,k_{x},\bar{y})$ follows from Eq. (\ref%
{2}) by using $|Y|$ in the RHS of Eq.(\ref{2}). As $\rho
^{a}(\omega ,k_{x},0)=0$ for antisymmetric EMPs it follows from
Eq.(\ref{2}) that
\begin{equation}
\sum_{n=0}^{\infty }\rho _{n}^{a}(\omega ,k_{x})=0.  \label{3}
\end{equation}%
For $\bar{y}<0$, $\rho ^{a}(\omega ,k_{x},\bar{y})$ is defined by Eq.(\ref{2}%
) and the odd-parity property.

Finally, the symmetric and antisymmetric modes are defined by
\begin{eqnarray}
(\omega -k_{x}v_{g})&&\rho _{m}^{s,a}(\omega ,k_{x})-\sum_{n=0}^{\infty
}[Sr_{mn}^{s,a}(k_{x})  \nonumber \\
&&+S^{\prime }g_{mn}^{s,a}(k_{x})]\rho _{n}^{s,a}(\omega ,k_{x})=0,  \label{4}
\end{eqnarray}%
where
$S=(2/\epsilon )(k_{x}\sigma _{yx}^{0}-ik_{x}^{2}\tilde{%
\sigma}_{yy})$, $S^{\prime }=-2i\tilde{\sigma}_{yy}/\epsilon \ell _{T}^{2}$,
\begin{eqnarray}
&&r_{mn}^{s,a}(k_{x})=\ell _{T}\int_{0}^{\infty }dx\ e^{-x}L_{m}(x)  \nonumber \\
&&\times \int_{0}^{\infty }dx^{\prime }\{[K_{0}(|k_{x}|\ell _{T}|x-x^{\prime
}|)  \nonumber \\
&&+\beta K_{0}(|k_{x}|\ell _{T}\sqrt{(x-x^{\prime })^{2}+(2d/\ell _{T})^{2}})]
\nonumber \\
&&\pm \lbrack K_{0}(|k_{x}|\ell _{T}|x+x^{\prime }|)  \nonumber \\
&&+\beta K_{0}(|k_{x}|\ell _{T}\sqrt{(x+x^{\prime })^{2}+(2d/\ell _{T})^{2}}%
)]\}  \nonumber \\
&&\times \tilde{R}_{0}(x^{\prime })e^{-x^{\prime }}L_{n}(x^{\prime }),
\label{5}
\end{eqnarray}%
and
\begin{eqnarray}
&&g_{mn}^{s,a}(k_{x})=|k_{x}|\ell _{T}^{2}\int_{0}^{\infty}dxe^{-x}\{e^{-x/2}L_{m}(x)/\cosh (x/2)  \nonumber \\
&&-\frac{m}{x}[L_{m}(x)-L_{m-1}(x)]\}  \nonumber \\ 
&&\times \int_{0}^{\infty }dx^{\prime }\{[\text{sign}\{x-x^{\prime
}\}K_{1}(|k_{x}|\ell _{T}|x-x^{\prime }|)  \nonumber \\ 
&&+\beta \frac{(x-x^{\prime })K_{1}(|k_{x}|\ell _{T}\sqrt{(x-x^{\prime
})^{2}+(2d/\ell _{T})^{2}})}{\sqrt{(x-x^{\prime })^{2}+(2d/\ell
_{T})^{2}}}] \nonumber \\ 
&&\pm [K_{1}(|k_{x}|\ell _{T}(x+x^{\prime }))
\nonumber \\
&& +\beta \frac{(x+x^{\prime })K_{1}(|k_{x}|\ell
_{T}\sqrt{(x+x^{\prime
})^{2}+(2d/\ell _{T})^{2}})}{\sqrt{(x+x^{\prime })^{2}+(2d/\ell _{T})^{2}}}%
]\}  \nonumber \\
&& \times \tilde{R}_{0}(x^{\prime })e^{-x^{\prime }}L_{n}(x^{\prime })+\delta
g_{mn}^{s,a}(k_{x}).  \label{6}
\end{eqnarray}%
Here sign$\{x\}=1$ for $x>0$ and sign$\{x\}=-1$ for $x<0$, $K_{1}(x)$ is the
modified Bessel functions. In addition, $\delta g_{mn}^{s}(k_{x})\equiv 0$
and
\begin{eqnarray}
&&\delta g_{mn}^{a}(k_{x})=2|k_{x}|\ell _{T}^{2}\int_{0}^{\infty }dx\
[K_{1}(|k_{x}|\ell _{T}x)  \nonumber \\
&&+\beta \frac{xK_{1}(|k_{x}|\ell _{T}\sqrt{x^{2}+(2d/\ell _{T})^{2}})}{\sqrt{%
x^{2}+(2d/\ell _{T})^{2}}}]\tilde{R}_{0}(x)e^{-x}L_{n}(x).  \label{7}
\end{eqnarray}%
Point out, for antisymmetric EMPs, due to Eq.(\ref{3}) the logarithmically divergent 
contributions from the terms Eq.(\ref{7}) are mutually cancelled,
after the summation over $n$ in Eq.(\ref{4}). 
Hereafter we drop $k_{x}$ in $r_{mn}^{s,a}(k_{x})$
and $g_{mn}^{s,a}(k_{x})$ to simplify the notation; notice that
$r_{mn}^{s(a)}\neq r_{nm}^{s(a)}$, $g_{mn}^{s(a)}\neq
g_{nm}^{s(a)}$.

In the solution of Eqs. (\ref{4})-(\ref{7}), we are taking the
long-wavelength limit $|k_{x}|\ell _{T}\ll 1$. As a result, for $\beta =0$
we can use the approximation $K_{0}(|k_{x}|\ell _{T}x)\approx \ln
(2/|k_{x}|\ell _{T})-\gamma -\ln (x)$ and $K_{1}(|k_{x}|\ell _{T}x)\approx
(|k_{x}|\ell _{T}x)^{-1}$, where $\gamma $ is the Euler constant. However,
for $\beta \neq 0$ when $(2d/\ell _{T})^{2}\gg 1$, such that $%
2|k_{x}|d\gtrsim 1$, we will use exact expressions for pertinent modified
Bessel functions. We also will denote $\overline{\omega }=\omega -k_{x}v_{g}$%
, where typically $v_{g}\ll \omega /k_{x}$ for the most fast EMP modes,
i.e., the renormalized monopole EMPs. So the main contribution to their
phase velocity, $\omega /k_{x}$, is given by $\overline{\omega }/k_{x}$. The
latter contribution is due to the electron-electron interactions induced by
wave excitation. 

For definiteness, if otherwise it is not stated, we consider 2DES in the
homogeneous sample, \textit{i.e.}, without the gate or the air half-plane.

\section{EMPs at $\protect\nu =1, 2$}

Point out, we can neglect the intermode coupling by omitting in
Eq.(\ref{4}) all nondiagonal coefficients $r_{m,n}^{s,a}$ and
$g_{m,n}^{s,a}$, \textit{i.e.}, with $m\neq n$. Then from Eqs.(\ref{2}), (\ref{4}) it follow
pure symmetric modes (monopole, only $\rho
_{0}^{s}\neq 0$, quadrupole, only $\rho _{1}^{s}\neq 0$,
\textit{etc}.) or pure antisymmetric modes (dipole, only $\rho
_{0,1}^{a}\neq 0$, \textit{etc}.). However, coupling among the
\textquotedblleft neighboring\textquotedblright\ pure modes
\cite{parity} due to nondiagonal, $m\neq n$, coefficients
$r_{m,n}^{s,a}$ and $g_{m,n}^{s,a}$ typically is strong and
especially for \textit{very strong dissipation}. So the coupling
between different pure modes should be taken into account. This
leads to renormalized modes, which there are true EMPs. In
particular, the modification of dissipation strength from
\textit{the weak dissipation} to \textit{very strong dissipation}
should strongly effect the characteristics of renormalized EMPs.
The latter ones we call also as EMPs. For the assumed
long-wavelength limit $k_{x}\ell _{T}\ll 1$
we define the weak, strong and very strong
dissipative regimes more precisely by the conditions $\eta _{T}\ll
1$, $\ln (1/k_{x}\ell _{T})\gg \eta _{T}
\agt 1$ and $\eta_{T} \agt 
\ln (1/k_{x}\ell _{T})\gg 1$,
respectively; $\eta _{T}=\xi /k_{x}\ell _{T}$ where $%
\xi =\tilde{\sigma}_{yy}/(\ell _{T}\sigma _{yx}^{0})$, notice,
$|S^{\prime }|/S=\eta_{T}$. Because the magnetic field is strong
it follows that $\xi \ll 1$. However, as $k_{x}\ell _{T}\ll 1$ the
regime of very strong dissipation can be easily
achieved if $v_{g}\geq s$.

Point out, similar regimes (conditions) hold for a sample with
the metal gate; only changing $\ln (1/k_{x}\ell _{T})$ by
$\ln(d/\ell _{T})$,  if  $\ln(d/\ell _{T})$ is large. Notice that here 
$Im \overline{\omega }(k_{x})\equiv Im\omega (k_{x})$, since only real $%
k_{x} $ is used in the present study. Furthermore, due to assumed
$k_{x}\ell _{T}\ll 1 $ and $\xi \ll 1$ typically we have $S \approx Re S $.

\subsection{Helical edge magnetoplasmon for very strong dissipation regime}

Here we are now looking for weakly damped EMPs, $Re\overline{\omega
}(k_{x})/[-Im\omega (k_{x})]\gg 1 $. From a detailed
analysis of the pure modes (monopole, quadrupole, dipole and
octupole) spectra and their change when only the coupling with one
neighboring pure mode is considered,
we conclude that merely the renormalized monopole
mode can be the candidate. Indeed, both for the strong and the weak
dissipation regimes, it was found\cite{balev00} that when only the
$n=0$ term is considered in Eq. (\ref{2}) together with the $m=0$
equation for $\rho _{0}^{s}(\omega ,k_{x})$ in Eq.(\ref{4})
$(-Im\overline{\omega }(k_{x}))/|S^{\prime }|$ $\approx
0.12$ is almost six times larger than the same ratio for the
partly renormalized monopole EMP in the limit of $k_{x}\ell
_{T}\rightarrow 0$ when two terms ($n=0,1$) in Eq. (\ref{2}) are
taken into account. In the latter case two equations from Eq.
(\ref{4}), with $m=0,1$, for $\rho _{0,1}^{s}(\omega ,k_{x})$ are
considered. This shows that for the monopole EMP, even for the weak
or the strong dissipation, both the one term and the two terms
approximations do not show good convergence in calculation of
$Im\overline{\omega }$, while for
$Re\overline{\omega }$ both these approximations already
demonstrate a very good convergence. Moreover, for very strong
dissipation, our present treatment of the general dispersion
relation obtained within two terms approximation (for two coupled
symmetric EMPs) as\cite{balev00}
\begin{eqnarray}
\omega _{\pm }^{s}&=&k_{x}v_{g}+\frac{1}{2}[S(r_{00}^{s}+r_{11}^{s})+S^{\prime
}(g_{00}^{s}+g_{11}^{s})]  \nonumber \\
&&\pm \frac{1}{2}\{[S(r_{00}^{s}-r_{11}^{s})+S^{\prime
}(g_{00}^{s}-g_{11}^{s})]^{2}  \nonumber \\
&&+4(Sr_{01}^{s}+S^{\prime }g_{01}^{s})(Sr_{10}^{s}+S^{\prime
}g_{10}^{s})\}^{1/2},  \label{8}
\end{eqnarray}%
shows that $-Im\overline{\omega }(k_{x})/|S^{\prime }|$ can be even
less than $0.02$ for $k_{x}\ll 1/\ell _{T}$. In addition, it changes sign
(\textit{e.g.}, at $1/k_{x}\ell _{T}\approx 150$ for $\xi =0.1$) and then
its modulus quickly tends to a small value, $\approx 1.5\times 10^{-2}$, as $%
k_{x}$ further decreases. It is clear that obviously unphysical
positive values of $Im\omega $ are due to lack of precision
of the two-term expansion for the \textquotedblleft
monopole\textquotedblright\ EMP at very strong dissipation regime.

\begin{figure}[ht]
\includegraphics*[width=1.0\linewidth]{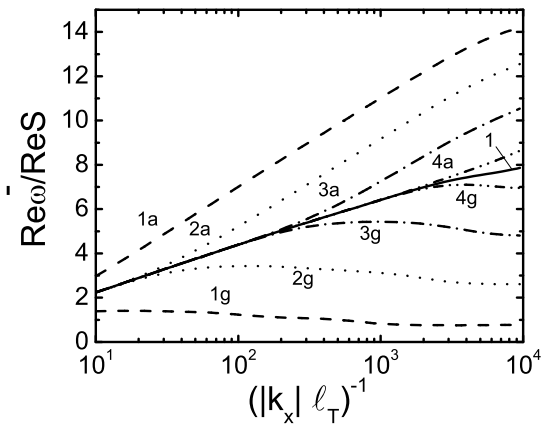}
\caption{\label{fig1}Dimensionless shifted phase velocity of the helical edge magnetoplasmon, 
$Re\protect\overline{\protect\omega }/Re S=(\protect\pi \protect\hbar \protect\epsilon/e^{2})[Re\protect\omega
/k_{x}-v_{g}]$, as a function of $1/k_{x}\ell _{T}$ for
$\protect\nu =1$, $\protect\xi =0.1$ in the regime of very strong
dissipation. The solid curve (1) corresponds to a homogeneous sample.
The lower (upper) dashed /1g(1a)/, dotted /2g(2a)/, dot-dashed /3g(3a)/, and dot-dot-dashed /4g(4a)/
curves correspond to a sample with the gate (air half-space) at
$d=10^{-5}$, $10^{-4}$, $10^{-3}$ and $10^{-2}$cm, respectively.}
\end{figure}

Point out that, in principle, above analytical considerations give
a chance for existence, even at $\eta_{T} \gg 1$, of a
\textit{window of transparency} for a \textquotedblleft
renormalized  monopole\textquotedblright\ EMP; \textit{i.e.}, a
finite range of $k_{x}$ in which the condition of weak damping can
be satisfied. Now we will show the possibility of a window of
transparency at very strong dissipation regime. The results
presented in all figures are calculated by keeping seven terms,
$n=0$,...,$6$, in the Eq.(\ref{2}) and
using seven equations, $m=0$,...,$6$, from Eq.(\ref{4}), that leads to a $%
7\times 7$ linear system of homogeneous equations. The the condition of
a nontrivial solution gives the dispersion relations for seven renormalized EMP
modes: monopole, dipole, quadrupole, etc. The
seventh term approximation shows very good convergence for the
properties of the renormalized monopole, dipole, and quadrupole
EMP modes. However, the only EMP mode that we will show in Figs.
1-3 is the renormalized monopole EMP.

In Fig. \ref{fig1} we plot
the shifted dimensionless phase velocity, $Re\overline{\omega }/Re S$,
of the renormalized monopole EMP as a
function of $k_{x}$ for 
$\nu =1$ and $\xi =0.1.$ 
Here, for the model confining potential Eq. (\ref{A0}), 
it is used $B=4.1$T, $T=9$K, $\omega _{c}/\Omega =30$, 
$E_{F0}=\hbar \omega _{c}$; then $\ell _{T}\approx 4.2\times 10^{-6}$ cm, 
$2\ell _{T}/\ell _{0}\approx 6.6$, the 2DES density 
$n_{0}\approx 10^{11} cm^{-2}$ and $v_{g}>s$. 
Point out, self-consistent many-body results
of Ref. \cite{irina2001} for $v_{g}$ give $\xi =0.1$ and
$v_{g}\approx 4.6\times 10^{5}$cm/sec $>s$, \textit{e.g.},
for $B\approx 15.7$T, $T\approx 10$K; $2\ell
_{T}/\ell _{0}\approx 4.0$, $n_{0}\approx 3.8\times
10^{11} cm^{-2}$.

From Fig. \ref{fig1} it is seen that the shifted phase velocity decreases
(increases) when the sample with metal gate (air half-space) is
considered as compared with that in the homogeneous sample. Notice
that this phase velocity is induced by electron-electron
interactions and it is as well dependent on the dissipative wave
processes.

\begin{figure}[ht]
\includegraphics*[width=1.0\linewidth]{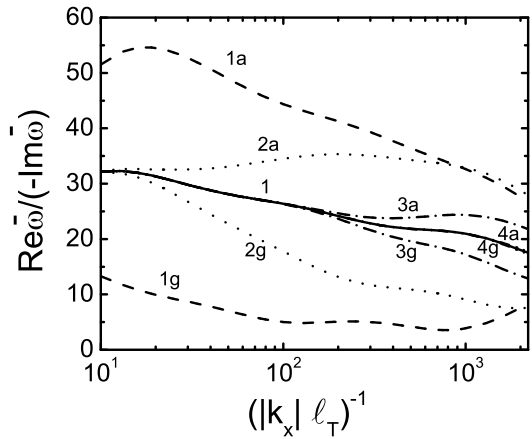}
\caption{\label{fig2}Plot of $Re\overline{\protect\omega }/(-Im\overline{%
\protect\omega })=Re(\protect\omega -k_{x}v_{g})/(-Im\protect%
\omega )$ as a function of $1/k_{x}\ell _{T}$ for the helical edge magnetoplasmon in the
very strong dissipation regime. The parameters are the same as in Fig. \ref{fig1}.
The solid and lower (upper) dashed, dotted, dot-dashed, dot-dot-dashed
curves correspond to pertinent curves in Fig. \ref{fig1}. For the homogeneous sample
(the solid curve) there is a clear window of transparency, where $Re%
\overline{\protect\omega }/(-Im\overline{\protect\omega
})\gg 1$, due to electron-electron interactions. It is seen that
when the gate is very close to the 2DES, the window of
transparency is essentially suppressed since the electron-electron
interaction become very weak due to screening effects.}
\end{figure}

In Fig. \ref{fig2}, we plot $Re\overline{\omega }/(-Im\overline{\omega }%
)$ of the renormalized monopole mode in the actual region of $k_{x}$, for the
same samples as in Fig. 1. We observe a very clear window of transparency in
the wide range $2\times 10^{3}\gtrsim (k_{x}\ell _{T})^{-1}\gtrsim 10^{2}$
to this mode, in which the very strong dissipation regime is achieved. In
this $k_{x}$-range the renormalized monopole EMP (helical edge magnetoplasmon) is weakly
damped, 
despite of very strong dissipation. Our analysis
shows that all other EMPs (both symmetric and antisymmetric) are quite
strongly damped with $-Im\overline{\omega }/Re\overline{\omega
}\gtrsim \eta _{T}\gg 1$. From Fig. \ref{fig2} we see that: (i) the gate screening
decreases $Re\overline{\omega }/(-Im\overline{\omega })$, but
the helical edge magnetoplasmon mode is still weakly damped for $d\gg \ell _{T}$;(ii) the effect
of the air half-space in the sample increases $Re\omega /(-Im%
\omega )$ as compared with the homogeneous sample.

\begin{figure}[ht]
\includegraphics*[width=1.0\linewidth]{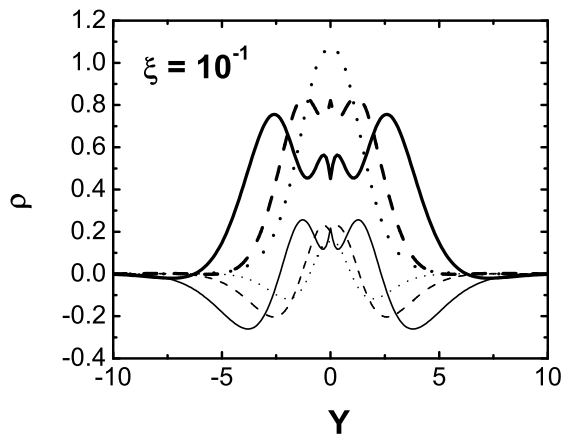}
\caption{\label{fig3} Dimensionless charge density profile $\protect\rho (Y)$ of helical edge
magnetoplasmon for conditions of Figs. 1, 2 for the homogeneous sample; 
$\protect\nu =1$, $\protect\xi =0.1$. The solid, dashed and dotted curves correspond
to $\protect\eta _{T}=10^{2}$, $10$ and $1$, respectively. The thick and the thin
curves correspond to the wave phase $\protect\phi=(Re\protect\omega \; t-k_{x}x)$ equal to
$2\protect\pi N_{k}$ and $\protect\pi /2+2\protect\pi N_{n}$, respectively, with any integer $N_{k}$, $N_{n}$;
in particular, for $N_{k}=N_{n}$.}
\end{figure}

In order to evaluate the spatial structure of the helical edge magnetoplasmon, we
solve,
for a given $k_{x}$ and correspondent complex dispersion $\overline{\omega }=%
\overline{\omega }(k_{x})$, the linear system of $(N-1)\times (N-1)$
independent inhomogeneous equations for the $(N-1)$ variables defined by $%
\rho _{j}(\overline{\omega }(k_{x}),k_{x})/\rho _{0}(\overline{\omega }%
(k_{x}),k_{x})$, $j=1,..,N-1$. Using these values in Eq.(\ref{2}) we
calculate the normalized charge density profile $\tilde{\rho}^{s}(\overline{%
\omega },k_{x},Y)=4\ell _{T}\rho ^{s}(\overline{\omega },k_{x},\bar{y})/\rho
_{0}^{s}(\overline{\omega },k_{x})$. Since this function is complex, we plot
in Fig. \ref{fig3} the density profile $\rho (Y)\equiv \rho ^{s}(Y,k_{x})=Re[%
\tilde{\rho}^{s}(\overline{\omega },k_{x},|Y|)\times \exp (i\phi )]$ for two
sets of the wave phase $\phi$. 
Thick and thin curves correspond to $\phi
=2\pi N_{k}$ and $\pi /2+2\pi N_{n}$, where $N_{k}$, $N_{n}$ can be any integer. 
In Fig. 3 the density profiles 
are calculated from the $6\times 6$ systems of linear inhomogeneous equations: 
for the solid curves $Re\overline{\omega }/ReS=0.06372$ and $Im\overline{\omega }/Re S=-0.00327$, for the
dashed curves $Re\overline{\omega }/Re S=0.44132$ and $Im\overline{\omega }/Re S=-0.01712$,
and for the dotted curves (plotted for comparison as the condition of very strong dissipation 
is not satisfied for them, while it holds for the solid and the dashed curves) 
$Re\overline{\omega }/Re S =2.236$ and
$Im\overline{\omega }/Re S=-0.0621$.   It can be seen that
for $\phi =2\pi N_{k,n}$ the density profiles have large
monopole contributions that lead to $\int_{-\infty }^{\infty }dY\rho
(Y)\neq 0$. On the other hand, for $\phi =\pi /2+2\pi N_{k,n}$ the
monopole contribution to $\rho (Y)$ is nullified so $\int_{-\infty
}^{\infty }dY\rho (Y)=0$. Fig. \ref{fig3} clearly show a strong presence of 
the multipole contributions, in particular, the quadrupole one. 

Even though in Figs. 1-3 we have used $\xi =0.1$, similar qualitatively results
are obtained for $10^{-1} \agt \xi \agt 10^{-3}$ which corresponds to
realistic values $v_{g}>s$.

\subsection{Fundamental mode in the weak dissipation regime}

The treatment of the damping rate for the fundamental (or monopole) EMP in
the weak dissipation regime, $\eta _{T}\ll 1$, was given in 
Ref. \onlinecite{balev00}. However, the result was obtained by keeping only two
terms in the expansion of Eq.(\ref{2}). As it was pointed out
above this approximation is rather rough for determination of the
damping rate. Now we will show, in particular, that three terms expansion,
in Eq.(\ref{2}), well approximates the exact result for
$Im\omega $ of the fundamental EMP of $n=0$ LL.

\begin{figure}[ht]
\includegraphics*[width=1.0\linewidth]{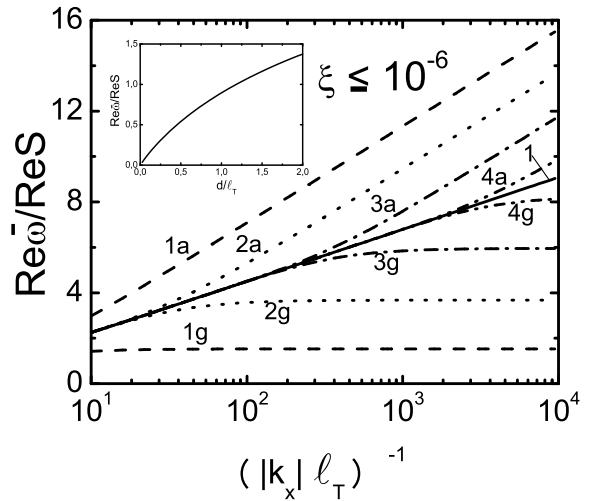}
\caption{\label{fig4} Dimensionless shifted phase velocity, $Re\overline{\protect%
\omega }/Re S$, of the fundamental EMP of the $N=0$ LL, for $\protect%
\xi \leq 10^{-6}$, within the long-wavelength region where $10^{-1}\geq
\protect\eta _{T}\geq 10^{-4}$ or smaller. As in Fig. \ref{fig1}, 
the solid curve (1) corresponds to a homogeneous sample;
the lower (upper) dashed /1g(1a)/, dotted /2g(2a)/, dot-dashed /3g(3a)/, and dot-dot-dashed /4g(4a)/
curves correspond to a sample with the gate (air half-space) at
$d=10^{-5}$, $10^{-4}$, $10^{-3}$ and $10^{-2}$cm, respectively.
The inset shows $Re\overline{\protect\omega }/Re S$ as a function of 
$d/\ell _{T}$ for a $k_{x}$ range where the
dispersion is purely acoustic in the case of the sample with a gate. Only
for $d/\ell _{T}\lesssim 10^{-1}$ this dependence becomes linear and it is
well approximated by $Re\overline{\protect\omega }/k_{x}\approx (%
\protect\pi \protect\sigma _{yx}^{0}/\protect\epsilon )(d/\ell _{T})$.}
\end{figure}

In Fig. \ref{fig4} we depict the dimensionless shifted phase velocity, $Re\overline{%
\omega }/Re S$, of the fundamental EMP for $\xi \leq 10^{-6},$ $\nu =1$
and $\ell _{T}\approx 4.2\times 10^{-6}$cm, in the long-wavelength
interval given by $10^{4}\geq (k_{x}\ell _{T})^{-1}\geq 10$, where $%
10^{-1}\geq \eta _{T}\geq 10^{-4}$ (or smaller, if $\eta_{T} <
10^{-6}$) in such a way that the condition of weak dissipation is
well satisfied. 
For a given $d$ it is seen that the curves for a sample with gate or air
are almost equal to the solid curve if $(k_{x}\ell _{T})^{-1}\leq d/2\ell
_{T}$. However, for $(k_{x}\ell _{T})^{-1}>d/2\ell _{T}$ the curves for a
sample with gate or air begin to depart from the solid curve. In particular,
for the sample with gate, the dispersion tends to be acoustic-like, 
$Re\omega /k_{x}=$const. The inset in the Fig. \ref{fig4} shows $Re\overline{%
\omega }/Re S$ as a function of $d/\ell _{T}$ for a $k_{x}$ range
where the mode dispersion is purely acoustic in the sample with a gate. We
point out that the distance dependence of $Re\overline{\omega }/Re S$ is 
linear only for $d/\ell _{T}\leq 10^{-1}$. Using only one term in
the expansion of Eq.(\ref{2}), we obtain $Re\overline{\omega }%
/k_{x}\approx 2\pi \lbrack 2ln(2)-1](\sigma _{yx}^{0}/\epsilon )(d/\ell
_{T}) $ for $d/\ell _{T}\ll 1$. When more terms are taken into account the
shifted phase velocity increases slightly and it is better approximated by $Re%
\overline{\omega }/k_{x}\approx (\pi \sigma _{yx}^{0}/\epsilon
)(d/\ell _{T}) $. These asymptotic values are in qualitative
agreement with estimates given in Ref. \cite{zhitenev95} for the
phase velocity of the EMP in a sample with the gate very close to
the 2DES. Note that in GaAs based samples $\xi \ll 10^{-3}$ can be
achieved only for $v_{g}<s$. 
Here, for the model confining potential Eq. (\ref{A0}), it is used: $B=4.1$T,
$T=4.5$K, $\omega _{c}/\Omega =60$, $E_{F0}=\hbar \omega
_{c}$ and we have $v_{g}<s$.

\begin{figure}[ht]
\includegraphics*[width=1.0\linewidth]{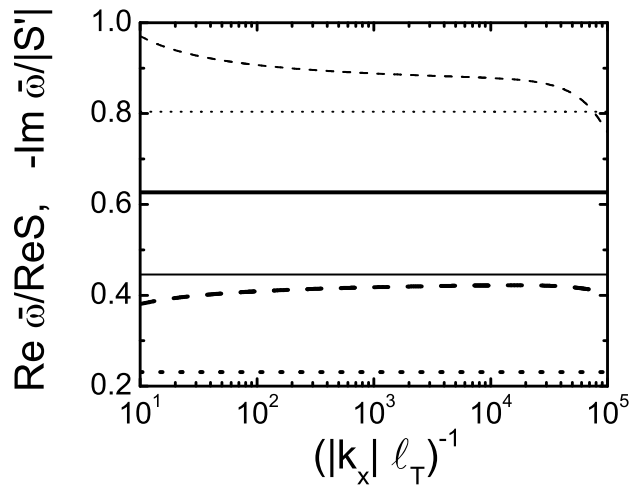}
\caption{\label{fig5} Dimensionless shifted phase velocity $Re\overline{\protect%
\omega }/ Re S$ (thick curves) and dimensionless imaginary part of frequency
(thin curves) for dipole, quadrupole and octupole $\protect\nu =1$ EMPs
plotted for a weak dissipation ($\protect\xi =10^{-6}$) by the solid,
the dashed, and the dotted curves, respectively. These curves are almost the same for
other values of $\protect\xi $ within the weak dissipation regime.}
\end{figure}

In Fig. \ref{fig5} at $\nu =1$ for $\xi=10^{-6}$, i.e., within the weak dissipation regime, $\eta _{T}\ll
1$, we present by the thin curves the dimensionless damping rate, $-Im\overline{\omega
}/|S^{\prime }|$, and by the thick curves the dimensionless shifted phase velocity, $Re\overline{\omega }%
/Re S$. The solid, the dashed, and the dotted curves correspond to the dipole, the quadrupole, and the 
octupole EMPs. These results are obtained from the solution of the $7\times 7$
system of linear homogeneous equations for symmetric and
antisymmetric EMPs, when a very good convergence is achieved for
the dispersion relations. From Fig. \ref{fig5} it is seen that only the
shifted phase velocity and the damping rate of the quadrupole EMP (the
second symmetric mode) are dependent on $k_{x}$. Those for the
dipole and octupole EMPs (first two antisymmetric modes) are
independent of $k_{x}$. We see that the ratio
$Re\overline{\omega}/Im\overline{\omega }$ for the
dipole EMP is larger than that for the quadrupole EMP. The latter 
is larger than that for the octupole EMP. The same 
holds for the shifted phase velocity. As expected, the dipole EMP
is the most weakly damped after the fundamental EMP of $n=0$ LL in
the weak dissipation regime at $\nu =1$. Although Fig. \ref{fig5} is pertinent
to the the homogeneous sample, the effect of the gate or air 
half-space on these higher-order EMP modes is
negligible for any $k_{x}$ if $d/\ell _{T}\gg 1$.
\begin{figure}[ht]
\includegraphics*[width=1.0\linewidth]{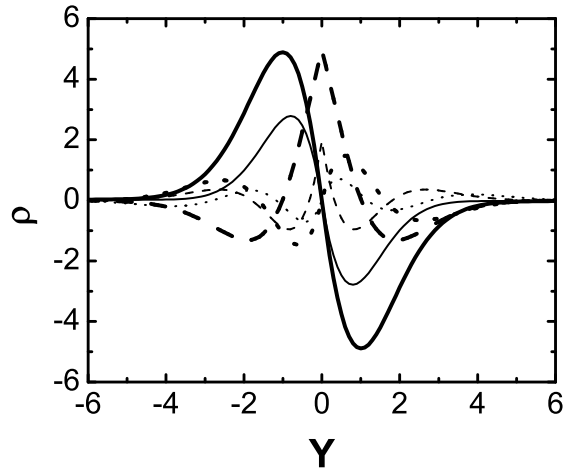}
\caption{\label{fig6} Dimensionless charge density profile $\protect\rho (Y)$ for
conditions taken from Fig. \ref{fig5}; in particular, $\protect\xi=10^{-6}$, $\protect\eta%
_{T}=10^{-2}$, and $k_{x}\ell_{T}=10^{-4}$. The thick and thin curves correspond to the wave phase $%
\protect\phi=2\protect\pi N_{k}$ and $\protect\pi /2+2\protect\pi N_{n}$, respectively, for any
integer $N_{k}$, $N_{n}$; the thin curves plot $10 \times \protect\rho (Y)$. The solid, dashed and
dotted curves correspond to the dipole, the quadrupole, and the octupole EMPs,
respectively.}
\end{figure}

In Fig. \ref{fig6}, we show the charge density profile $\rho (Y)\equiv \rho
_{s,a}(Y,k_{x})=Re[\tilde{\rho}_{s,a}(\overline{\omega }%
,k_{x},Y)\times \exp (i\phi )]$ of the higher-order EMPs for two sets of
the wave phase $\phi $ and $\xi =10^{-6}$. The charge density for dipole,
quadrupole and octupole EMPs is indicated by thick (thin) solid, dashed and
dotted curves, respectively, for $\eta _{T}=10^{-2}$. The curves were
calculated from the $6\times 6$ system of linear inhomogeneous equations.
The thick and thin curves correspond to $\phi =2\pi N_{k}$ and $\pi /2+2\pi
N_{n}$, respectively. The thin curves are $10\times \rho $. In Fig. \ref{fig6}, we
have $\overline{\omega }=0.626S+0.446S^{\prime }$ for the solid curves, $%
\overline{\omega }=0.422S+0.878S^{\prime }$ for the dashed curves and $%
\overline{\omega }=0.231S+0.804S^{\prime }$ for the dotted curves. We
observe that the density profiles of the dipole EMP is not modified
qualitatively by varying the wave phase $\phi $.

\section{Fundamental EMPs at $\protect\nu =4$}

Now, assuming the weak dissipation at the $\nu =4$ QHE regime, we
will treat fundamental EMPs omitting dissipation.
For definiteness, only the model confining potential Eq. (\ref{A0}) is considered. 
Then, instead of Eq. (\ref{1}), the integral equation is given as

\begin{eqnarray}
&&\sum_{n=0}^{1}(\omega -k_{x}v_{gn})\rho _{n}(\omega ,k_{x},y) 
-\widetilde{S}\sum_{n=0}^{1}R_{0n}\left( y\right) \nonumber \\
&&\times \int_{-\infty }^{\infty
}dy^{\prime }[\beta K_{0}(|k_{x}|\sqrt{(y-y^{\prime })^{2}+4d^{2}})  \nonumber \\
&&+K_{0}(|k_{x}||y-y^{\prime }|)] \sum_{m=0}^{1}\rho _{m}(\omega ,k_{x},y^{\prime })=0,  
\label{9}
\end{eqnarray}%
where $\widetilde{S}=(2\widetilde{\sigma }_{yx}^{0}/\epsilon )k_{x}$, $%
\widetilde{\sigma }_{yx}^{0}=e^{2}/\pi \hbar $, $R_{0n}\left( y\right)
=R_{0n}\left( \overline{y}_{n}\right) =\left( 4\ell _{Tn}\right) ^{-1}\cosh
^{-2}\left( \overline{y}_{n}/2\ell _{Tn}\right) $, $\ell _{Tn}=\ell
_{0}k_{B}T/\hbar v_{gn}$, and $v_{g}=\hbar \Omega ^{2}k_{re}^{\left(
n\right) }/m^{\ast }\omega _{c}^{2}$ is the group velocity of the edge
states of the $n$-th LL; $\overline{y}_{n}=y-y_{re}^{\left( n\right) }$, $%
y_{re}^{\left( n\right) }=\ell _{0}^{2}k_{re}^{\left( n\right) }$, $n=0,1$.
The Hall conductivity in the inner part of the channel is given by $\sigma
_{yx}^{0}=2\widetilde{\sigma }_{yx}^{0}=2e^{2}/\pi \hbar $. We assume that $%
2\ell _{Tn}/\ell _{0}\gg 1$ and the long-wavelength limit $\left\vert
k_{x}\right\vert \ell _{Tn}\ll 1$, $\left\vert k_{x}\right\vert \Delta
y_{01}\ll 1$, where $\Delta y_{01}=y_{re}^{\left( 0\right) }-y_{re}^{\left(
1\right) }$. In addition, it was assumed in Eq.(\ref{9}) that $\Delta
y_{01}/\ell _{T1}\gg 1$ and the charge density distortion of the $n$-th LL
is localized nearby its edge, $y_{re}^{\left( n\right) }.$ Note that 
the LLs spectra for our model Eq. (\ref{A0}) give that $v_{g1}<v_{g0}$ and,
hence, $\ell _{T1}>\ell _{T0}$. We can look for the solution of Eq.(\ref{9})
as
\begin{equation}
\rho \left( \omega ,k_{x},y\right) =\sum_{n=0}^{1}\rho _{n}\left( \omega
,k_{x},y\right) ,  \label{10}
\end{equation}%
where $\rho _{n}\left( \omega ,k_{x},y\right) $ are localized within a
region of length of the order of $\ell _{Tn}$ around the edge of the $n$-th
LL. To study fundamental EMPs of the $n=0,1$ LLs, $\rho _{n}\left(
\omega ,k_{x},y\right) \equiv \rho _{n}\left( \omega ,k_{x},\overline{y}%
_{n}\right) $ is well approximated by
\begin{equation}
\rho _{n}\left( \omega ,k_{x},\overline{y}_{n}\right) =R_{0n}\left(
\overline{y}_{n}\right) \rho _{n}\left( \omega ,k_{x}\right) .  \label{11}
\end{equation}%
Indeed, for $k_{x}\rightarrow 0$ the solution of Eq. (\ref{9}) given by Eqs.
(\ref{10}), (\ref{11}) is almost exact. Furthermore, as we have shown in
previous sections, similar approximations for the fundamental EMP of $n=0$
LL at $\nu =1\left( 2\right) $ yield very accurate results for the
dispersion relation and the charge density profile for all $k_{x}$. However, now the
symmetry of the problem with respect to $y_{re}^{\left( 0\right) }$ or $%
y_{re}^{\left( 1\right) }$ is broken due to the presence of the
inter-edge Coulomb interaction.

Neglecting by small overlapping between $\rho _{0}\left( \omega ,k_{x},\overline{y}_{0}\right)$
and $\rho_{1}\left( \omega ,k_{x},\overline{y}_{1}\right)$,
we multiply Eq.(\ref{9}) by $\widetilde{R}_{0n}^{-1}(Y_{n})$, where $%
\widetilde{R}_{0n}(Y_{n})=\exp (|Y_{n}|)\;R_{0n}(Y_{n})$, and integrate over
$Y_{n}=\overline{y}_{n}/\ell _{Tn}$ from $-\infty $ to $\infty $. Then,
taking into account Eqs. (\ref{10}) and (\ref{11}), the coupled system of
two equations for $\rho _{n}\equiv \rho _{n}\left( \omega ,k_{x}\right) $ is
given by

\begin{eqnarray}
&&\lbrack \omega -(k_{x}v_{g0}+\widetilde{S}A_{00}/2)]\rho _{0}-(\widetilde{S}%
A_{01}/2)\rho _{1}=0,  \nonumber \\
&&-(\widetilde{S}A_{01}/2)\rho _{0}+[\omega -(k_{x}v_{g1}+\widetilde{S}%
A_{11}/2)]\rho _{1}=0,  \label{12}
\end{eqnarray}%
where%
\begin{eqnarray}
A_{mn}\left( k_{x}\right) &=&\frac{1}{4}\int\limits_{-\infty }^{\infty
}dx \int\limits_{-\infty }^{\infty }dx^{\prime } \exp (-|x|) \cosh ^{-2}\left( x^{\prime }/2\right) \nonumber \\
&&\times [\beta K_{0}(|k_{x}|\ell _{Tn}\sqrt{(\widetilde{x}_{mn}\left( x\right)
-x^{\prime })^{2}+(2d/\ell _{Tn})^{2}})\nonumber \\
&&+K_{0}(|k_{x}|\ell _{Tn}|\widetilde{x}_{mn}\left( x\right) -x^{\prime }|)]  ,  
\label{13}
\end{eqnarray}%
with $\widetilde{x}_{mn}\left( x\right) =(\Delta y_{mn}/\ell
_{Tn})+(\ell _{Tm}/\ell _{Tn})x$. Point out that as the integral
term in Eq.(\ref{9}) is very weakly dependent on $y$ for the
assumed very small $k_{x}$, we still
can arrive to Eqs. (\ref{12}), (\ref{13}) even if the overlapping between $%
\rho _{0}\left( \omega ,k_{x},y\right) $ and $\rho _{1}\left(
\omega ,k_{x},y\right) $ is not very small.

First we will study EMPs (i) for the sample without the gate
($\beta =0$). Secondly, we will study EMPs (ii) for $d/\ell
_{Tn}\gg 1$ and (iii) for $d/\ell _{Tn}\ll 1$ in the case of the
sample with gate ($\beta =-1$), assuming that $|k_{x}|d\ll 1$. For
the case (i) ($\beta =0$) we have $ A_{01}=A_{10}\approx 2[\ln
(1/|k_{x}|\ell _{T1})+\ln (2)-\gamma +\ln (\ell _{T1}/\Delta
y_{01})]$, $A_{11}\approx 2[\ln (1/|k_{x}|\ell _{T1})+\ln
(2)-\gamma ]-0.25$, and $A_{00}\approx A_{11}+2\ln (\ell
_{T1}/\ell _{T0})$. Thus, despite the condition $\Delta
y_{01}/\ell _{T1}\gg 1$, the system of equations, Eq.(\ref{12}),
is strongly coupled by the long-range Coulomb interaction between
the edges of the LLs. If this inter-edge Coulomb coupling is
neglected, by setting $A_{01}=0$, the Eq.(\ref{12}) yields the
dispersion relations of the \textit{decoupled} fundamental EMPs of
the $n=0$ (for $\rho _{0}\neq 0$) and $n=1$ (for $\rho _{1}\neq
0$) LL's as

\begin{eqnarray}
\omega ^{(n)} &=&k_{x}v_{gn}+(2/\epsilon )\widetilde{\sigma }_{yx}^{0}k_{x}
\nonumber \\
&&\times \left[\ln \left(1/\left\vert k_{x}\right\vert \ell _{Tn}\right)-0.01\right].
\label{14}
\end{eqnarray}%
If we take into account the Coulomb coupling between these fundamental
EMP's, their spectra, as it follows from Eq.(\ref{12}), changes drastically.
The dispersion of the renormalized fundamental EMP of the $n=0$ LL, or the
fast fundamental EMP, becomes

\begin{eqnarray}
\omega _{+}^{\left( 01\right) } &=&k_{x}v_{01}+(4/\epsilon )
\widetilde{\sigma }_{yx}^{0}k_{x}  \nonumber \\
&&\times \lbrack \ln (1/\left\vert k_{x}\right\vert \ell _{c}^{+})+0.05],
\label{15}
\end{eqnarray}%
where $v_{01}=\left( v_{g0}+v_{g1}\right) /2$, and $\ell _{c}^{+}=\sqrt{%
\Delta y_{01}\sqrt{\ell _{T0}\ell _{T1}}}$ is the characteristic length that
determines the spatial dispersion of the fast mode. On turn, the
renormalized fundamental EMP of the $n=1$ LL, or the slow fundamental EMP,
becomes

\begin{eqnarray}
\omega _{-}^{(01)} &=&k_{x}v_{01}+(2/\epsilon )\widetilde{\sigma }%
_{yx}^{0}k_{x}  \nonumber \\
&&\times \left[ \ln \left(\Delta y_{01}/\sqrt{\ell _{T1}\ell _{T0}}\right)-0.12\right],
\label{16}
\end{eqnarray}
where the ratio of the characteristic lengths $\Delta y_{01}$ and
$\sqrt{\ell _{T1}\ell _{T0}}$ mainly determine the dispersion of
the slow mode. It can be seen that $\omega _{-}^{\left( 01\right)
}\left( k_{x}\right)$ becomes purely acoustic in contrast to
$\omega _{+}^{\left( 01\right) }\left( k_{x}\right) $. From Eqs.
(\ref{14})-(\ref{16}) we conclude that renormalization of the dispersion $%
\omega ^{(n)}$ due to the Coulomb coupling of the charge
fluctuations at the edges of $n=0,1$ LLs is strong as typically
$\omega _{+}^{(01)}/\omega _{-}^{(01)}\gg 1$ while $\omega
^{(0)}/\omega ^{(1)}\approx 1$. The phase velocities of both fast
and slow fundamental EMPs decrease with increasing $T $.

\begin{figure}[ht]
\includegraphics*[width=1.0\linewidth]{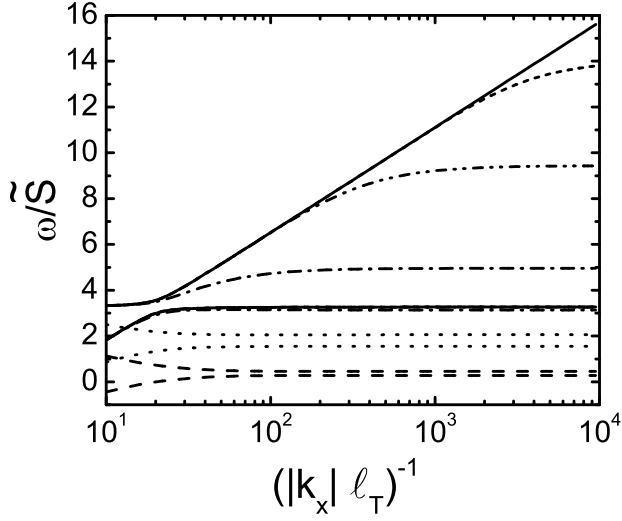}
\caption{\label{fig7} Dimensionless phase velocity of the fast and the slow fundamental EMPs, for $\protect\nu %
=4$, as a function of $(k_{x}\ell _{T})^{-1}$ calculated from Eq.(\protect\ref%
{12}); $\widetilde{S}=(2/\protect\epsilon )\widetilde{\protect\sigma }_{yx}^{0}k_{x}$. 
The solid curves correspond to the homogeneous sample. For
samples with the gate at $d=10^{-6}$cm ($\approx \Delta y_{01}/90$), 
$10^{-5}$cm ($\approx \Delta y_{01}/9$), $10^{-4}$cm ($\approx
\Delta y_{01}$), $10^{-3}$cm ($\approx 11\times \Delta y_{01}$), and 
$10^{-2}$cm ($\approx 110\times \Delta y_{01}$) from the 2DES the
plots are presented by the dashed, dotted, dash-dotted,
dash-dot-dotted and shorted-dash curves, respectively.}
\end{figure}

For the case (ii) ($\beta =-1$ and $d/\ell_{T_{n}} \gg 1$) we find
$A_{01}=A_{10}\approx 2\ln
(2d/\Delta y_{01})$, $A_{11}\approx 2\ln (2d/\ell _{T1})-0.25$, and $%
A_{00}\approx 2\ln (2d/\ell _{T0})-0.25$. In this case the
dispersion of the fast fundamental EMP it follows from Eq.
(\ref{12}) as

\begin{equation}
\omega _{+}^{\left( 01\right) }=k_{x}v_{01}+(4/\epsilon )\widetilde{\sigma }%
_{yx}^{0}k_{x}[\ln (2d/\ell _{c}^{+})-0.06],  \label{17}
\end{equation}%
while the slow fundamental EMP is still given by Eq.(\ref{16}).

Finally, in the case (iii) ($\beta =-1$ and $d/\ell_{T_{n}} \ll
1$), the dispersion relation for the renormalized fundamental EMPs
of $n=0,1$ LLs is given by

\begin{equation}
\omega _{n}^{\left( 01\right) }=k_{x}v_{gn}+\frac{2\pi }{\epsilon }\frac{d}{%
\ell _{Tn}}\widetilde{\sigma }_{yx}^{0}k_{x}[2\ln (2)-1].  \label{18}
\end{equation}%
As expected, $\omega _{0}^{\left( 01\right) }$ is in agreement with the
result for $\nu =1, 2$, obtained in Sec. IIIB, since the fundamental EMP
of $n=0$ LL is totally decoupled from the fundamental EMP of $n=1$ LL.

In Fig. \ref{fig7} the dimensionless phase velocity, $\omega /\tilde{S}$,
for the renormalized fundamental EMPs of the $n=0$ LL and the
$n=1$ LL is plotted as
a function of $(k_{x}\ell _{T})^{-1}$, $\ell _{T}\equiv \ell _{T0},$ for $%
\nu =4$ from Eq.(\ref{12}). 
For each sample, the top curve corresponds to the
fast mode and the bottom curve represents the slow mode. For the
fast EMP mode only the samples with the gate can show
acoustic-like dispersion, while for the slow EMP mode in all
samples there is a wide region with the acoustic dispersion. 
Notice, for
$d \alt 10^{-6}$cm Eq.(\ref{17}) well approximates
dispersion relations for the fast fundamental EMP. Here, for a
GaAs-based heterostructure, it is assumed that $B=4.1$T,
$\omega _{c}/\Omega
\simeq 100$, $E_{F0}=2\hbar \omega _{c}$, $T=2.66$K; then we have $%
v_{g0}\approx 2.35\times 10^{5}$cm/sec $<s$, $v_{g1}\approx
1.36\times 10^{5}$cm/sec $<s$, $\ell _{T1}/\ell _{T0}=\sqrt{3}$ with $%
\ell _{T1}\approx 0.41\times 10^{-5}$cm and $2\ell
_{T0}/\ell _{0}\approx 3.8$. In addition, it follows that $\Delta
y_{01}\approx 9.3\times 10^{-5}$cm and $\Delta y_{01}/\ell
_{T1}\approx 22.5$.

\begin{figure}[ht]
\includegraphics*[width=1.0\linewidth]{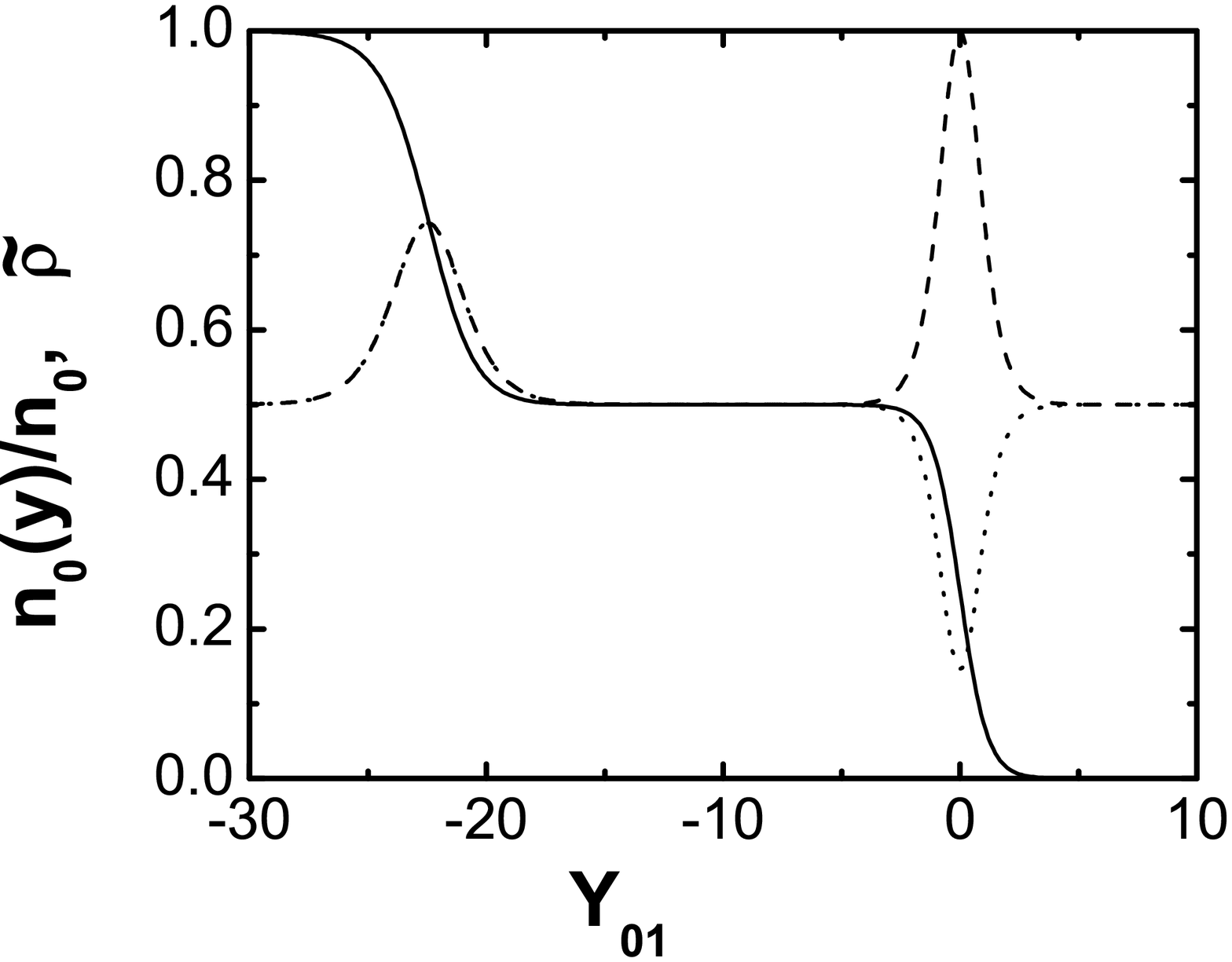}
\caption{\label{fig8} The unperturbed electron density $n_{0}\left( y\right) $,
normalized to the bulk value $n_{0}$, and the dimensionless charge density
profile $\widetilde{\protect\rho }\left( \protect\omega ,k_{x},Y_{01}\right)
\equiv 4\ell _{T1}\protect\rho \left( \protect\omega ,k_{x},Y_{01}\right) /%
\protect\rho _{1}$, where $Y_{01}=\overline{y}_{0}/\ell _{T1}$, of the
renormalized fundamental EMPs for $\left( \left\vert k_{x}\right\vert \ell
_{T1}\right) ^{-1}=10^{2}$. The solid curve corresponds to $n_{0}\left(
y\right) $, the dashed curve corresponds to the fast fundamental EMP, $%
\protect\omega _{+}^{\left( 01\right) }/\tilde{S}=6.52$, and the
dotted curve correspond to the slow fundamental EMP,
$\protect\omega _{-}^{\left( 01\right) }/\tilde{S}=3.25$. Here the
same parameters are used as for the solid curves in Fig. \ref{fig7}. The
dashed and the dotted curves plot $\widetilde{\protect\rho }/4.1$,
shifted upward on 0.5. }
\end{figure}

Substituting Eqs. (\ref{15}) and (\ref{16}) in Eqs. (\ref{10})-(\ref{12})
the charge density profiles of the fast and the slow fundamental EMPs, in
the absence of gate in the sample, are given by

\begin{eqnarray}
\widetilde{\rho }\left( \omega ,k_{x},Y_{01}\right) &=&\left( \ell _{T1}/\ell
_{T0}\right) \cosh ^{-2}\left( \left( \ell _{T1}/\ell _{T0}\right)
Y_{01}/2\right) \widetilde{\rho }_{0}  \nonumber \\
&&+\cosh ^{-2}\left( \left( \Delta y_{01}/2\ell _{T1}\right) +Y_{01}/2\right) ,
\label{19}
\end{eqnarray}%
where $\widetilde{\rho }\left( \omega ,k_{x},Y_{01}\right) =4\ell _{T1}\rho
\left( \omega ,k_{x},Y_{01}\right) /\rho _{1}$, $\widetilde{\rho }_{0}=\rho
_{0}/\rho _{1}$, and $Y_{01}=\overline{y}_{0}/\ell _{T1}$. In Fig. \ref{fig8} for the
same conditions that for the solid curves in Fig. \ref{fig7} we plot the charge
density profile for the fast fundamental EMP, $\frac{1}{4.1} \widetilde{%
\rho }(\omega _{+}^{(01)},k_{x},Y_{01})$, by the dashed curve and
for the slow fundamental EMP, $\frac{1}{4.1} \widetilde{\rho
}\left( \omega_{-}^{(01)},k_{x},Y_{01}\right) $, by the dotted
curve. 
In Fig. \ref{fig8} it is assumed that $y_{re}^{\left( 0\right)
}=0$; $y_{re}^{\left( 1\right) }/\ell _{T1}=-\Delta y_{01}/\ell
_{T1}\approx -22.5$ and $y_{r}/\ell _{T1}\approx -53.2$. 

\section{CONCLUDING REMARKS}

In present microscopic treatment of EMPs we, e.g., similar to Refs. \cite{volkov88,aleiner94}, have neglected the diffusion 
current. Here this approximation is well justified, except for the charge density profiles $\rho(Y)$ of the symmetric modes 
within a very narrow region, $|Y| \ll \ell_{T}$, as it is seen in Figs. 3, 6.
Where small finite cusps (notice, in Ref. \cite{volkov88} the cusp of density is infinite, without the diffusion current)
are present. Indeed, in the absence of the diffusion current, for symmetric modes our equations warrant
at $Y=0$ the continuity of $\rho(Y)$, however, not of its first derivative, $d\rho(Y)/dY$;
inclusion of the diffusion current will make $d\rho(Y)/dY$ continuous at $Y=0$ as well.
So, cf. with Ref. \cite{volkov88}, when  the diffusion current is taken into account it will only 
smooth out these small cusps, without any essential effect as on the EMPs spectra so on the charge density profile.

In present study,
it is used simple analytical model, Eq. (\ref{A0}), for the
confining potential $V^{\prime}(y)$ (in addition, for some figures we present
also the estimations that include the effect of many-body interactions\cite{irina2001}).
This analytical model reproduces quite well the confining
potential of a wide channel calculated numerically in the
Hartree approximation \cite{balev00,balev2001,zozoulenko}.
Moreover, present treatment can be easily extended
for the confining potentials of different forms if they are smooth on
the scale of $\ell_{0}$ and the not-too-low temperatures conditions
are satisfied.
In particular,
the group velocity of the edge states $v_{g}$ renormalized by
self-consistent exchange-correlation effects 
can be finite even if the Hartree approximation group velocity $v_{g}^{H}=0$
\cite{irina2001}. The latter condition corresponds to the LL
flattening and the formation of the compressible strip \cite{glazman92}, cf. with
\cite{brey94,ando93} and \cite{zozoulenko}. Notice that with
exchange and correlation
interactions included within density functional theory
approximation (rather crude for 2DES in the quantum Hall regime, see,
e.g., \cite{balev2001})
also it is found that for realistic parameters the
exchange interaction can completely suppress the formation of
compressible strips \cite{zozoulenko}.

So the applicability of our
present results may have substantially wider temperatures range
than $30$K $\agt T \agt 10$K. Furthermore, we can
speculate that treated here model and helical edge magnetoplasmon at not-too-low
temperatures can be relevant to EMPs interference observed
recently \cite{kukushkin04,kukushkin05,dorozhkin05} for
temperatures $\leq 80$K  as it, in particular, needs
weakly damped EMPs.\cite{mikhailov06}

\begin{acknowledgments}
We thank Nelson Studart for many useful discussions. This work was supported by
Brazilian FAPEAM  (Funda\c{c}\~{a}o de Amparo \`{a} Pesquisa do Estado do
Amazonas) Grants; this work of O. G. B. was also supported by Brazilian CNPq
Grant.
\end{acknowledgments}


\begin{thebibliography}{10}

\bibitem{allen83} S. J. Allen, H. L. Stormer, and J. C. M. Hwang, Phys. Rev.
B \textbf{28}, 4875 (1983).

\bibitem{mast85} D. B. Mast, A. J. Dahm, and A. L. Fetter, Phys. Rev. Lett.
\textbf{54}, 1706 (1985).

\bibitem{glattli85} D. C. Glattli, E. Y. Andrei, G. Deville, J. Poitrenaud,
and F. I. B. Williams, Phys. Rev. Lett. \textbf{54}, 1710 (1985).

\bibitem{galchenkov86} S. A. Govorkov, M. I. Reznikov, A. P. Senichkin, and
V. I. Talyanskii, Pis'ma Zh. Eksp. Teor. Fiz. \textbf{44}, 380 (1986) [JETP
Lett. \textbf{44} , 487 (1986)]; V. A. Volkov, D. V. Galchenkov, L. A.
Galchenkov, I. M. Grodnenskii, O. R. Matov, and S. A. Mikhailov, Pis'ma Zh.
Eksp. Teor. Fiz. \textbf{44}, 510 (1986) [JETP Lett. \textbf{44} , 655
(1986)].

\bibitem{talyanskii89} V. I. Talyanskii, I. E. Batov, B. K. Medvedev, J. P.
Kotthaus, M. Wassermeier, A. Wixforth, J. Weimann, W. Schlapp, and H. Nikel,
Pis'ma Zh. Eksp. Teor. Fiz. \textbf{50}, 196 (1989) [JETP Lett. \textbf{50},
221 (1989)].

\bibitem{wassermeier90} M. Wassermeier, J. Oshinowo, J. P. Kotthaus, A. H.
MacDonald, C. T. Foxon, and J. J. Harris, Phys. Rev. B \textbf{41}, 10287
(1990).

\bibitem{grodnensky91} I. Grodnensky, D. Heitmann, and K. von Klitzing,
Phys. Rev. Lett. \textbf{67},1019 (1991); Surface Science \textbf{263}, 467
(1992).

\bibitem{ashoori92} R. C. Ashoori, H. L. Stormer, L. N. Pfeiffer, K. W.
Baldwin, and K. West, Phys. Rev. B \textbf{45}, 3894 (1992).

\bibitem{talyanskii92} V. I. Talyanskii, A. V. Polisski, D. D. Arnone, M.
Pepper, C. G. Smith, D. A. Ritchie, J. E. Frost, and G. A. C. Jones, Phys.
Rev. B \textbf{46}, 12427 (1992).

\bibitem{zhitenev93} N. B. Zhitenev, R. J. Haug, K. v. Klitzing, and K.
Eberl, Phys. Rev. Lett. \textbf{71}, 2292 (1993); Phys. Rev. B \textbf{49},
7809 (1994).

\bibitem{talyanskii95} V. I. Talyanskii, D. R. Mace, M. Y. Simmons, M.
Pepper, A. C. Churchill, J. E. F. Frost, D. A. Ritchie and G. A. C. Jones,
J. Phys. Condens. Matter \textbf{7}, L435 (1995).

\bibitem{ernst96} G. Ernst, R. J. Haug, J. Kuhl, K. von Klitzing, and K.
Eberl, Phys. Rev. Lett. \textbf{77}, 4245 (1996).

\bibitem{deviatov97} E. V. Deviatov, V. T. Dolgopolov, F. I. B. Williams, B.
Jager, A. Lorke, J. P. Kotthaus, A. C. Gossard, Appl. Phys. Lett. \textbf{71}%
, 3655 (1997).

\bibitem{balaban97} N. Q. Balaban, U. Meirav, H. Shtrikman, and V. Umansky,
Phys. Rev. B \textbf{55}, R13397 (1997); N. Q. Balaban, U. Meirav,
and I. Bar-Joseph, Phys. Rev. Lett. \textbf{81}, 4967 (1998).

\bibitem{sukhodub04} G. Sukhodub, F. Hohls, and R. J. Haug, Phys. Rev. Lett.
\textbf{93}, 196801 (2004).

\bibitem{kukushkin04} I. V. Kukushkin, M. Yu. Akimov, J. H. Smet, S. A.
Mikhailov, K. von Klitzing, I. L. Aleiner, and V. I. Falko, Phys.
Rev. Lett. \textbf{92}, 236803 (2004).

\bibitem{kukushkin05} I. V. Kukushkin, S. A. Mikhailov, J. H. Smet, and K.
v. Klitzing, Appl. Phys. Lett. \textbf{86}, 044101 (2005).

\bibitem{dorozhkin05} P. S. Dorozhkin, S. V. Tovstonog, S. A. Mikhailov, I.
V. Kukushkin, J. H. Smet, and K. v. Klitzing, Appl. Phys. Lett. \textbf{87}%
, 092107 (2005).

\bibitem{kirichek95} O. I. Kirichek, P. K. H. Sommerfeld, Yu. P. Monarkha,
P. J. M. Peters, Yu. Z. Kovdrya, P. P. Steijaert, R. W. van der Heijden, and
A. T. A. M. de Waele, Phys. Rev. Lett. \textbf{74}, 1190 (1995).

\bibitem{wen91} X. G. Wen, Phys. Rev. B \textbf{43}, 11025 (1991).

\bibitem{melikidze04} A. Melikidze and Kun Yang, Phys. Rev. B \textbf{70},
161312(R) (2004).

\bibitem{papa05} Wei-Cheng Lee, N. A. Sinitsyn, Emiliano Papa, and A. H.
MacDonald, Phys. Rev. B \textbf{72}, 121304(R) (2005).

\bibitem{fetter86} A. L. Fetter, Phys. Rev. B \textbf{33}, 3717 (1986).

\bibitem{volkov88} V. A. Volkov and S. A. Mikhailov, Zh. Eksp. Teor. Fiz.
\textbf{94}, 217 (1988) [Sov. Phys. JETP \textbf{67}, 1639 (1988)]; in
\textit{Modern Problems in Condensed Matter Sciences,} edited by V. M.
Agranovich and A. A. Maradudin (North-Holland, Amsterdam, 1991), Vol. 27.2,
Ch. 15, p. 885.


\bibitem{stone91} M. Stone, Ann. Phys. (NY) \textbf{207}, 38 (1991); M.
Stone, H. W. Wyld, and R. L. Schult, Phys. Rev. B \textbf{45}, 14156 (1992).

\bibitem{aleiner94} I. L. Aleiner and L. I. Glazman, Phys. Rev. Lett.
\textbf{72}, 2935 (1994).

\bibitem{chamon94} C. de Chamon and X. G. Wen, Phys. Rev. B \textbf{49},
8227 (1994).

\bibitem{giovanazzi94} S. Giovanazzi, L. Pitaevskii, and S. Stringari,
Phys. Rev. Lett. \textbf{72}, 3230 (1994).

\bibitem{han97} J. H. Han and D. J. Thouless, Phys. Rev. B \textbf{55},
R1926 (1997).

\bibitem{zulicke97} U. Zulicke, R. Bluhm, V. A. Kostelecky, and A. H.
MacDonald, Phys. Rev. B \textbf{55}, 9800 (1997).

\bibitem{balev97} O. G. Balev and P. Vasilopoulos, Phys. Rev. B \textbf{56},
13252 (1997).

\bibitem{balev98} O. G. Balev and P. Vasilopoulos, Phys. Rev. Lett. \textbf{%
81}, 1481 (1998); O. G. Balev, P. Vasilopoulos, and Nelson Studart, J.
Phys.: Condens. Matt. \textbf{11}, 5143 (1999).

\bibitem{balev99} O. G. Balev and P. Vasilopoulos, Phys. Rev. B \textbf{59},
2807 (1999).

\bibitem{balev00} O. G. Balev and Nelson Studart, Phys. Rev. B \textbf{61},
2703 (2000).

\bibitem{hansson00} T. H. Hansson and S. Viefers, Phys. Rev. B \textbf{61},
7553 (2000).

\bibitem{johnson03} M. D. Johnson and G. Vignale, Phys. Rev. B \textbf{67},
205332 (2003).


\bibitem{mikhailov06} S. A. Mikhailov, Appl. Phys. Lett. \textbf{89}, 042109
(2006).

\bibitem{zhitenev95} N. B. Zhitenev, R. J. Haug, K. v. Klitzing, and K.
Eberl, Phys. Rev. B \textbf{52}, 11277 (1995).

\bibitem{balev93} O. G. Balev and P. Vasilopoulos, Phys. Rev. B \textbf{47},
16410 (1993);  O. G. Balev, P. Vasilopoulos, and E. V. Mozdor,
Phys. Rev. B \textbf{50}, 8706 (1994);  O. G. Balev and P.
Vasilopoulos,  Phys. Rev. B \textbf{50}, 8727 (1994).

\bibitem{irina2001} I. O. Baleva, Nelson Studart, and O. G. Balev, Phys.
Rev. B \textbf{65}, 073305 (2002).

\bibitem{wave} Here $A(\omega,k_{x},y)$ is the wave amplitude of any wave physical value involved in
an EMP, e.g.: of the electron charge density, of the electric potential, any component of the electric field or 
of the current density.   

\bibitem{parity} E.g., for pure monopole mode most close neighboring pure mode
there is quadrupole one as coupling it is possible only among pure
modes of the same parity.\cite{balev00}


\bibitem{balev2001} O. G. Balev and Nelson Studart,
Phys. Rev. B \textbf{64}, 115309 (2001).

\bibitem{zozoulenko} S. Ihnatsenka and I. V. Zozoulenko,
Phys. Rev. B \textbf{73}, 075331 (2006);
Phys. Rev. B \textbf{73}, 155314 (2006);
Phys. Rev. B \textbf{78}, 035340 (2008).

\bibitem{glazman92} D. B. Chklovskii, B. I. Shklovskii, and L. I. Glazman,
Phys. Rev. B \textbf{46}, 4026 (1992).


\bibitem{brey94} L. Brey, J. J. Palacios, and C. Tejedor,
Phys. Rev. B \textbf{47}, 13884 (1993).

\bibitem{ando93} T. Suzuki and Tsuneya Ando, J. Phys. Soc. Jpn.
\textbf{62}, 2986 (1993); Physica B \textbf{201}, 345 (1994).




\end{thebibliography}

\end{document}